\documentclass{article}
\usepackage{spconf,amsmath,graphicx}

\usepackage{enumitem}
\setlist{nosep, leftmargin=14pt}

\usepackage{mwe} 

\usepackage{booktabs}
\usepackage{multirow}
\usepackage{bm}
\usepackage{subfig}
\usepackage{float}
\usepackage{amsfonts}

\title{A Benchmark Analysis of Graph and Non-Graph Methods for Caenorhabditis elegans Neuron Classification}
\name{Jingqi Lu$^{\dagger}$ \qquad Keqi Han$^{\dagger}$ \qquad Yun Wang$^{\star}$ \qquad Lu Mi$^{\ddagger}$ \qquad Carl Yang$^{\dagger}$ }
\address{$^{\dagger}$ Department of Computer Science, Emory University, USA\\
        $^{\star}$ Department of Biomedical Informatics, Emory University, USA\\
        $^{\ddagger}$ College of AI, Tsinghua University, China}
\begin{document}
%
\maketitle
\begin{abstract}
This study establishes a benchmark for \textit{Caenorhabditis elegans} neuron classification, comparing four graph methods (GCN, GraphSAGE, GAT, GraphTransformer) against four non-graph methods (Logistic Regression, MLP, LOLCAT, NeuPRINT). Using the functional connectome, we classified Sensory, Interneuron, and Motor neurons based on Spatial, Connection, and Neuronal Activity features. Results show that attention-based GNNs significantly outperform baselines on the Spatial and Connection features. The Neuronal Activity features yielded poor performance, likely due to the low temporal resolution of the underlying neuronal activity data. Our benchmark validates the use of GNNs and highlights that Spatial and Connection features are key predictors for \textit{Caenorhabditis elegans} neuron classes. Code is available at: https://github.com/JingqiLuu/neuronclf-gnn-benchmark.
\end{abstract}
\section{Introduction}
\label{sec:intro}

Neuron classification is a fundamental problem in neuroscience and is essential for understanding the organization and function of neural circuits~\cite{White1986}. In \textit{Caenorhabditis elegans} (\textit{C. elegans}), neurons are categorized into three functional classes: Sensory, Interneurons, and Motor~\cite{White1986,Randi}. Owing to its compact and fully mapped nervous system, \textit{C. elegans} serves as an ideal platform for studying this problem. Recently, Randi et al.~\cite{Randi} constructed a neural signal propagation atlas that captures functional signal dynamics not predictable from anatomy alone. This atlas provides both functional connectome information and neuronal activity data, enabling data-driven neuron classification and deeper insights into information flow in neural circuits.~\cite{Chalfie1985}.

Traditionally, neuron identification relied on static features such as morphology or molecular markers~\cite{GABA,NeuroPAL,Taylor2021}. With the advent of large-scale neural recording technologies, researchers have increasingly explored classification methods based on neuronal activity time series~\cite{LOLCAT, NeuPRINT, NEMO}. Recent deep learning models, such as LOLCAT~\cite{LOLCAT} and NeuPRINT~\cite{NeuPRINT}, have shown promise in classifying neurons by leveraging individual neuronal activity. In parallel, modeling the brain as a complex network has driven the rapid adoption of Graph Neural Networks (GNNs) in macroscopic brain network analysis, especially for tasks like disease diagnosis or biomarker identification~\cite{BrainGB, Kimisbi2024, han2025rethinking}. Despite these advances, the integration of activity-based representations and graph-based learning frameworks for neuron classification at the microscale, especially in the \textit{C. elegans} functional connectome, remains largely unexplored.

This study establishes a benchmark for neuron classification in \textit{C. elegans}. We systematically evaluate graph methods (GNNs) against non-graph methods (e.g., LOLCAT~\cite{LOLCAT}, NeuPRINT~\cite{NeuPRINT}) using the functional connectome~\cite{Randi} and three distinct feature types: Spatial, Connection, and Neuronal Activity features. Specifically, our study aims to (1) assess the relative performance of graph-based and non-graph-based approaches, (2) identify the most informative feature type for neuron classification, and (3) compare the efficacy of different GNN architectures.

\section{Method}
\label{sec:method}

\subsection{Dataset}
The data for this study are drawn from the openly accessible \textit{C. elegans} functional connectome published by Randi et al.~\cite{Randi}. This connectome is modeled as a directed weighted graph $G = (V, E, W)$. The node set $V$ consists of the 300 neurons in the \textit{C. elegans} nervous system. The edge set $E$ contains 8,703 directed edges, representing functional signal propagation, while $W$ is the corresponding weighted adjacency matrix where $W_{ij}$ denotes the functional connection strength from neuron $i$ to $j$. For the neuron classification task, we adopt the dataset's intrinsic functional categorization as ground truth labels, which classifies neurons into three distinct classes: Sensory (90 neurons), Interneuron (93 neurons), and Motor (117 neurons)~\cite{Randi}. The dataset also provides the raw data for feature construction (detailed in Section~\ref{sec:features}), including the neuronal activity data for each neuron (2Hz sampling rate) and their spatial coordinates.

\subsection{Feature Construction}
\label{sec:features}
Based on best practices in brain network analysis~\cite{BrainGB} and research on neuronal activity patterns~\cite{LOLCAT, NeuPRINT}, we constructed and evaluated the following three types of features:

\begin{itemize}
\item \textbf{Spatial Feature}: The feature for each neuron is a 3-dimensional vector $f_{\text{spatial}} = (x, y, z)$, representing its standardized coordinates in physical space.
\item \textbf{Connection Feature}: Each neuron's feature $f_{\text{conn}} = W_{i, :}$ is its row vector in the adjacency matrix $W$, capturing its functional connectivity with all other neurons~\cite{BrainGB}.
\item \textbf{Neuronal Activity Feature}: This feature is constructed following LOLCAT~\cite{LOLCAT}. The continuous neuronal activity data for each neuron ($X_i(t)$) is first segmented into non-overlapping windows ($W_{i,k}$) of duration $T_w$. Within each window, we detect events via thresholding to calculate Interspike Intervals (ISIs). For each window $W_{i,k}$, these ISIs are summarized into an empirical distribution vector $\mathbf{d}_{i,k} \in \mathbb{R}^{D_{ISI}}$, where the value at index $\tau-1$ represents the count of ISIs equal to $\tau$ time steps (for $1 \le \tau \le D_{ISI}$). The final feature for neuron $i$ is a sequence $\mathcal{S}_i = (\mathbf{d}_{i,1}, ..., \mathbf{d}_{i,N_k})$. Sequences are padded to a uniform maximum length $L_{max}$ for batch processing, resulting in a feature tensor $F_{ISI} \in \mathbb{R}^{N \times L_{max} \times D_{ISI}}$, where $N$ denotes the number of neurons.
\end{itemize}

\begin{table*}[htbp]
\centering
\ninept 
\setlength{\tabcolsep}{5pt} 
\caption{Performance Comparison of Graph and Non-graph Methods for Neuron Classification Across Three Feature Sets. For each metric (column), the highest score is in \textbf{bold} and the second-highest score is \underline{underlined}.}
\label{result_table}
\begin{tabular}{cccccccc}
\toprule
\multirow{2}{*}{\textbf{Type}} & \multirow{2}{*}{\textbf{Model}} & \multicolumn{2}{c}{\textbf{Spatial Feature}} & \multicolumn{2}{c}{\textbf{Neuronal Activity Feature}} & \multicolumn{2}{c}{\textbf{Connection Feature}} \\
\cmidrule(lr){3-4} \cmidrule(lr){5-6} \cmidrule(lr){7-8}
 &  & \textbf{Accuracy} & \textbf{F1} & \textbf{Accuracy} & \textbf{F1} & \textbf{Accuracy} & \textbf{F1} \\
\midrule
\multirow{4}{*}{Non-graph method} & Logistic Regression & $56.33_{\pm 4.52}$ & $54.17_{\pm 4.38}$ & $36.59_{\pm 6.20}$ & $17.86_{\pm 6.11}$ & $54.67_{\pm 5.62}$ & $52.49_{\pm 6.56}$ \\
 & MLP & $56.67_{\pm 2.98}$ & $55.52_{\pm 3.69}$ & $40.00_{\pm 5.25}$ & $31.81_{\pm 4.93}$ & $54.67_{\pm 1.94}$ & $53.00_{\pm 1.75}$ \\
 & LOLCAT & $57.00_{\pm 2.67}$ & $55.86_{\pm 2.31}$ & $39.00_{\pm 5.83}$ & $25.11_{\pm 8.84}$ & $54.00_{\pm 2.26}$ & $52.78_{\pm 2.28}$ \\
 & NeuPRINT & $55.33_{\pm 9.33}$ & $50.87_{\pm 10.61}$ & $45.57_{\pm 8.69}$ & $37.19_{\pm 14.18}$ & $50.67_{\pm 4.90}$ & $42.51_{\pm 13.08}$ \\
\midrule
\multirow{4}{*}{Graph method} & GCN & $60.00_{\pm 6.65}$ & $54.38_{\pm 8.16}$ & $45.00_{\pm 8.50}$ & $40.53_{\pm 9.83}$ & $58.05_{\pm 7.14}$ & $44.87_{\pm 12.29}$ \\
 & GraphSAGE & $57.07_{\pm 4.52}$ & $44.09_{\pm 5.54}$ & $\bm{50.00_{\pm 10.21}}$ & $\bm{47.34_{\pm 10.11}}$ & $60.49_{\pm 4.73}$ & $52.37_{\pm 6.07}$ \\
 & GAT & \underline{$61.46_{\pm 2.84}$} & \underline{$57.12_{\pm 3.62}$} & $47.50_{\pm 10.07}$ & \underline{$46.03_{\pm 10.54}$} & \underline{$61.46_{\pm 9.56}$} & \underline{$55.08_{\pm 12.94}$} \\
 & GraphTransformer & $\bm{62.44_{\pm 3.96}}$ & $\bm{57.12_{\pm 4.46}}$ & \underline{$49.17_{\pm 9.65}$} & $45.87_{\pm 11.18}$ & $\bm{62.44_{\pm 6.28}}$ & $\bm{58.73_{\pm 5.77}}$ \\
\bottomrule
\end{tabular}
\end{table*}

\subsection{Model Architectures}Two main categories of models were compared: graph models and non-graph models~\cite{LOLCAT, NeuPRINT} that do not explicitly use the graph adjacency matrix. For each model-feature combination, optimal hyperparameters were determined via Bayesian optimization (detailed in Section~\ref{ImplementationDetails}). The resulting key parameters are summarized in their respective descriptions.

\noindent\textbf{Graph methods.} Four mainstream GNN architectures based on the message-passing paradigm were evaluated. These models iteratively update node representation by aggregating neighborhood features. In each layer $l$, node $v_i$'s representation $h_i^l$ is updated via message passing and updating.

In the message passing step, each node $v_i$ receives messages from all its neighbors. This process can be formally described by:
\begin{equation}
m_i^l = \sum_{j \in N_i} M^l(h_i^{l-1}, h_j^{l-1}, w_{ij}),
\end{equation}
where $N_i$ denotes the neighbor set of $v_i$, $h_j^{l-1}$ is the feature representation of a neighboring node $j$ from the previous layer, $w_{ij}$ represents the weight of the edge from node $j$ to $i$, and $M^l$ is the message function at layer $l$.

In the update step, the aggregated message $m_i^l$ is used to update the $v_i$'s representation from the previous layer:
\begin{equation}
h_i^l = U^l(h_i^{l-1}, m_i^l),
\end{equation}
where $U^l$ is the update function at layer $l$, which is typically a learnable linear transform with a non-linear activation. The GNN models, which differ in their definitions of the message ($M^l$) and update ($U^l$) functions, include:

\begin{itemize}
    \item \textbf{Graph Convolutional Network (GCN)}~\cite{GCN}: A classic GNN model that performs convolution by aggregating the weighted features of neighboring nodes. Optimized models used 2-6 layers and 30 to 160 hidden dimensions.
    \item \textbf{GraphSAGE}~\cite{GraphSAGE}: Employs neighborhood sampling and with various aggregation functions, demonstrating good scalability for large-scale graphs. Optimized models used 2-7 layers with hidden dimensions between 60 and 260.
    \item \textbf{Graph Attention Network (GAT)}~\cite{GAT}: Uses a multi-head attention mechanism to assign different importance weights to each neighbor for more flexible information aggregation. Optimized models used 2-5 layers, 40-190 hidden dimensions, and 1 to 8 attention heads.
    \item \textbf{GraphTransformer}~\cite{Graptaransformer}: Extends the Transformer architecture to graph-structured data by incorporating self-attention mechanisms that capture long-range dependencies and structural relationships between nodes. Optimized models used 2-4 layers, hidden dimensions between 50 and 200, and 1 to 8 attention heads.
\end{itemize}

\noindent\textbf{Non-Graph methods.} We also evaluated four models that do not explicitly use the graph adjacency matrix:
\begin{itemize}
    \item \textbf{Multi-Layer Perceptron (MLP)}: A standard feed-forward network. Optimized models used 2 to 6 layers with hidden dimensions varying between approximately 150 and 460.
    \item \textbf{Logistic Regression}: A multinomial linear classification model. Optimized models used regularization strength that C ranging from $10^{-4}$ to $10^{1}$ and L1 penalty type.
    \item \textbf{LOLCAT}~\cite{LOLCAT}: A supervised attention-based model designed to classify a neuron using only its isolated neuronal activity data. It operates by (1) dividing the full neuronal activity data into short "snippets", (2) computing the ISI distribution for each snippet, and (3) using a multi-head attention mechanism to learn which snippets are most informative for classification. Optimized models used 16-64 embedding/hidden dimensions and 4 attention heads.
    \item \textbf{NeuPRINT}~\cite{NeuPRINT}: A self-supervised framework learning a time-invariant representation for each neuron. It uses a Transformer model to predict a neuron's future activity based on its own past activity and permutation-invariant statistics of surrounding population activity. We use this framework to generate representation vectors (64-dim), which are then fed into an MLP classifier.
\end{itemize}

\subsection{Implementation Details}
\label{ImplementationDetails}
All model evaluations were performed using a stratified 5-fold Cross-Validation (CV) scheme for robustness. The dataset was split into five folds, with each fold serving once as the test set while the remaining four were used for training. Hyperparameters for each model-feature combination were optimized using the Optuna framework, an open-source tool designed for automating optimization processes~\cite{Optuna}. We employed Bayesian optimization with the Tree-structured Parzen Estimator (TPE) algorithm~\cite{TPE} as the sampler to efficiently search the hyperparameter space. Each model-feature combination ran 30 optimization trials. The optimization objective was to maximize the mean 5-fold CV validation accuracy. The hyperparameter search space included key parameters such as learning rate (log-uniform between $10^{-5}$ and $10^{-2}$), weight decay (log-uniform), hidden layer dimensions (uniform between 32 and 512), number of network layers (integer between 2 and 7), and dropout rate (uniform between 0.0 and 0.8). Model-specific parameters (e.g., attention heads for GAT) were also included in the search.

\begin{figure*}[t!] 
\centering

\begin{minipage}[t]{0.59\textwidth}
    \centering
    \subfloat[Example neuronal activity data for Sensory, Interneuron, and Motor neurons. \label{fig:neuron_activity}]
        {\includegraphics[width=\linewidth]{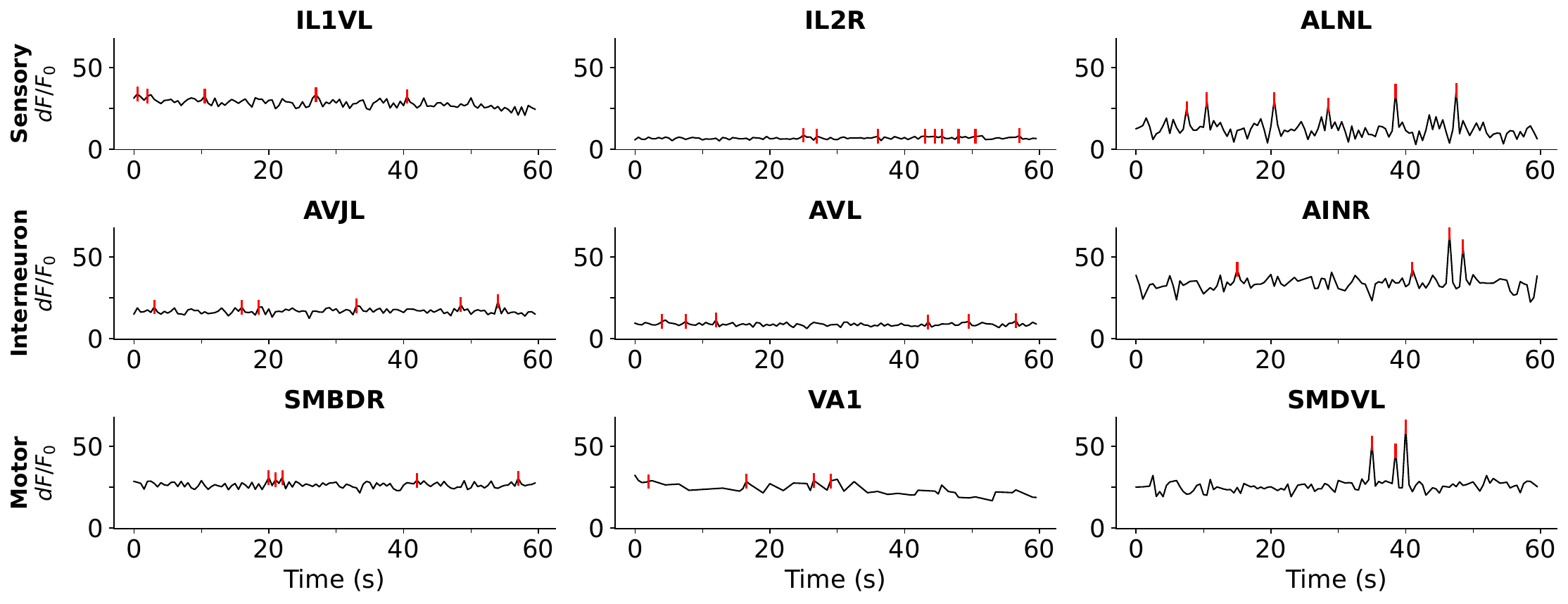}}
\end{minipage}
\begin{minipage}[t]{0.27\textwidth}
    \centering
    \subfloat[Cosine similarity. \label{fig:cosine_sim}]
        {\includegraphics[width=\linewidth]{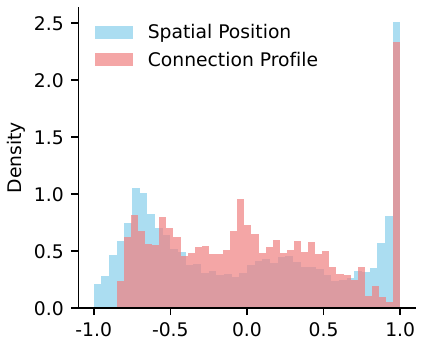}}
\end{minipage}

\begin{minipage}[t]{0.23\textwidth}
    \centering
    \subfloat[Head region distribution. \label{fig:atlas_overlay_head}]
        {\includegraphics[width=\linewidth]{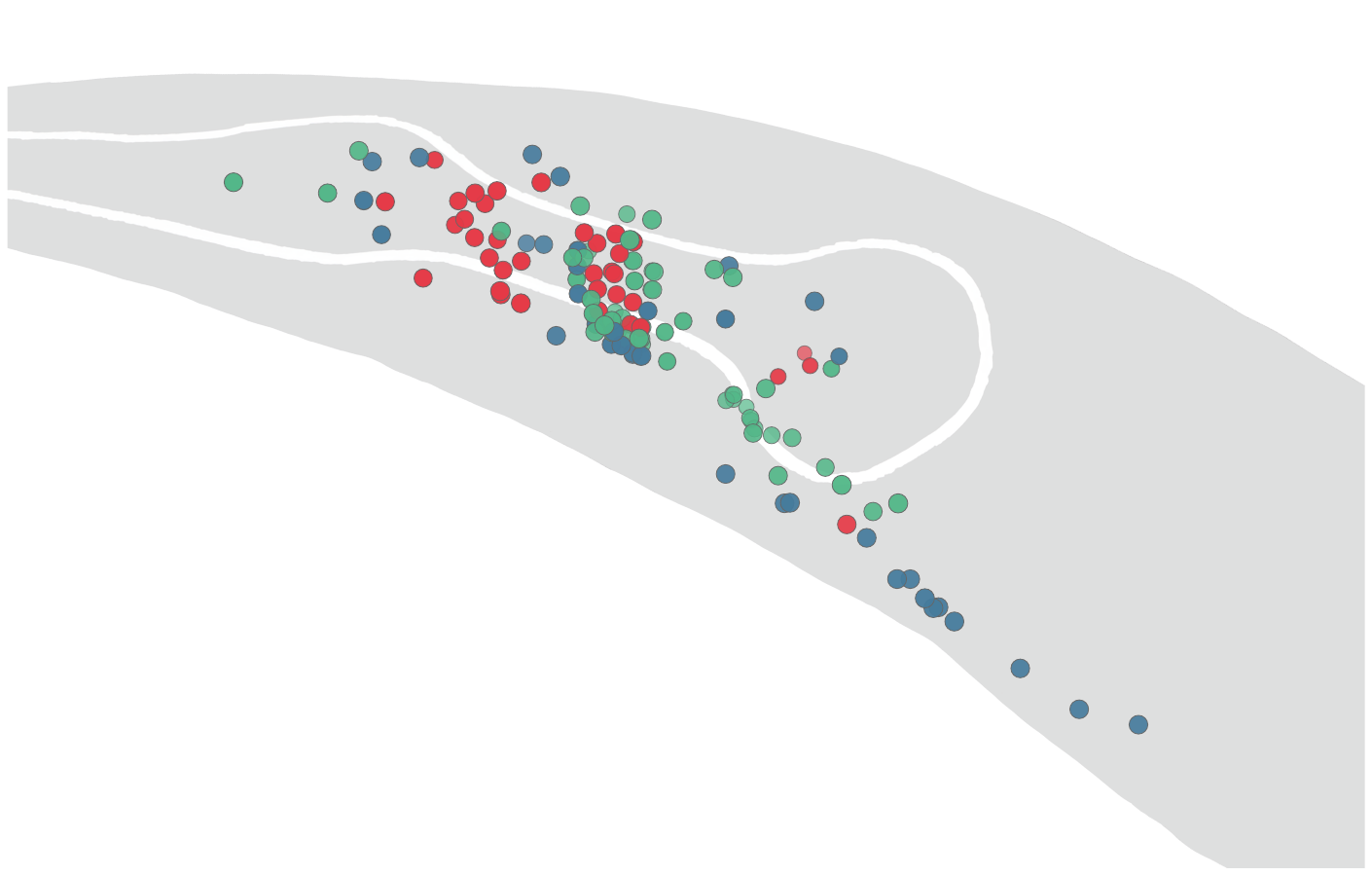}}
\end{minipage}
\begin{minipage}[t]{0.68\textwidth}
    \centering
    \subfloat[Spatial distribution of neuron classes. \label{fig:atlas_overlay}]
        {\includegraphics[width=\linewidth]{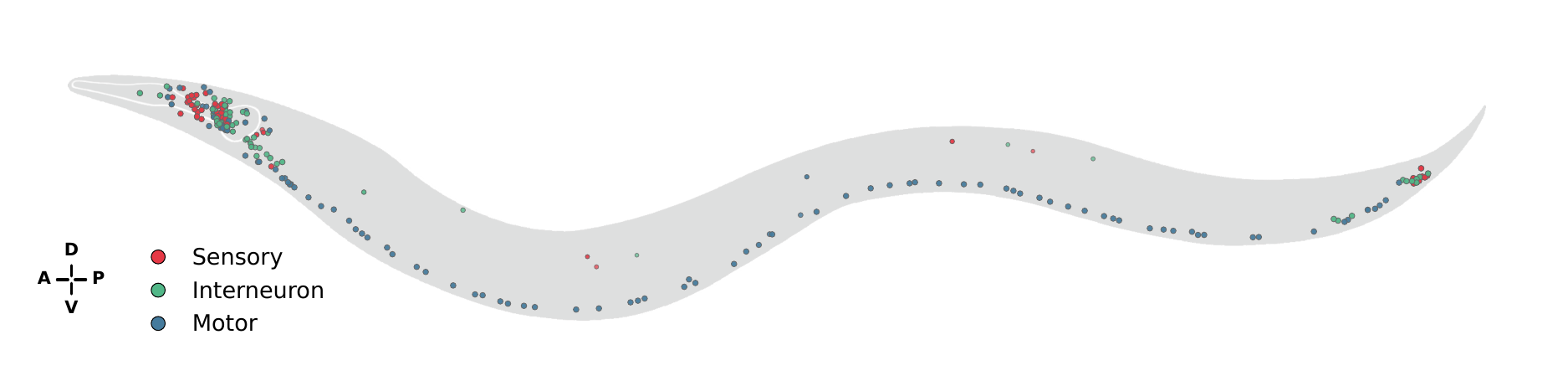}}
\end{minipage}

\caption{Visualization of \textit{C. elegans} neuron data features and class distributions. 
(a) Example neuronal activity data for representative Sensory, Interneuron, and Motor neurons. 
(b) Normalized probability density distributions of cosine similarity, comparing the Spatial and Connection features across all neuron pairs. 
(c) Magnified view of the spatial distribution of neuron classes in the head region. 
(d) Spatial distribution of the three neuron classes along the entire body.}
\label{fig:analysis}
\end{figure*}

\section{Results}
\label{sec:results}
We evaluated all models across the three feature sets. The reported results represent the mean accuracy and F1 score (± standard deviation) obtained from 5-fold cross-validation. Experimental results are summarized in Table~\ref{result_table}.

\subsection{Comparing Features}
\noindent\textbf{Spatial and Connection Features.} The Spatial and Connection features are effective for distinguishing the three major classes of \textit{C. elegans} neurons, with most models achieving accuracies around 55\%-62\%. This indicates that a neuron's functional class is closely related to both its connection patterns within the functional connectome and its spatial layout.

The effectiveness of Spatial feature stems from its strong class information: As shown in Fig.\ref{fig:atlas_overlay_head} and Fig.\ref{fig:atlas_overlay}, the \textit{C. elegans} nervous system is highly structured. Sensory neurons (red) and interneurons (green) are predominantly located in the head region, while motor neurons (blue) are primarily distributed along the ventral nerve cord.

\noindent\textbf{Spatial-Connectivity Feature Correlation. } An interesting finding is that the Spatial and Connection feature, although conceptually different, ultimately yielded very similar classification performance. To investigate this, we analyzed the intrinsic correlation between these two feature sets. This analysis involved constructing N×N similarity matrices for both feature sets using Cosine similarity. We then computed the Pearson correlation between the Spatial and Connection Feature similarity matrices. As illustrated in Fig.~\ref{fig:cosine_sim}, the similarity distributions of the two features exhibit similar morphologies. We found a significant positive correlation: the Pearson correlation coefficient for the matrices derived from Pearson similarity was 0.324 ($p < 0.001$). This indicates that neurons that are physically close in space also tend to have similar functional connectivity patterns. This moderate coupling explains their comparable classification outcomes.

\noindent\textbf{Neuronal Activity Feature. } In contrast, the Neuronal Activity feature yielded poor performance on the neuron classification task, all models generally achieve accuracies below 50\%. This suggests ISI are not distinct for these classes, or the low-sampling-rate (2Hz) data fails to capture them.

We attribute this poor performance primarily to the temporal resolution limitations of the dataset used in this study. As illustrated by the example raw neuronal activity data in Fig.~\ref{fig:neuron_activity}, the signals are smooth and lack the sharp transients typical of high-temporal-resolution recordings. Event detection performed on these smooth signals (indicated by red lines in the figure) reveals that many neurons produce a sparse number of events within the observation window. This low sampling rate limits the ability to accurately resolve fast neural events and may lead to the loss of high-frequency information; as noted in the LOLCAT study, a lower sampling rate impacts the quality of time-based features\cite{LOLCAT}. This signal smoothness and event sparsity directly impact models reliant on temporal patterns: it might obscure the local temporal patterns that LOLCAT seeks, as many windows contain too few events to form a meaningful ISI distribution. 

\subsection{Comparing Methods}
\noindent\textbf{Graph Methods. } Attention-based GNNs achieved the highest performance on the two most informative features. When using the Spatial feature, GraphTransformer achieves the highest accuracy (62.44\%) and a leading F1 score (57.12\%), with GAT also showing strong performance (61.46\%). When using the Connection Feature, GraphTransformer (62.44\%) and GAT (61.46\%) again perform best, significantly outperforming all non-graph models. This highlights the advantage of attention-based GNNs in capturing complex functional connectivity patterns, as they can adaptively focus on the most informative connections for neuron classification.

An important finding is how the graph structure's contribution varied by feature. On Spatial and Connection features, the performance gain of GNNs over the strongest non-graph baselines was moderate. This is likely because, as established in Section 3.1.2, these features are already intrinsically correlated with the graph topology; neurons physically close tend to have similar connections. Thus, explicitly adding the graph structure provides only a limited amount of new, non-redundant information. Conversely, on the Neuronal Activity feature, the graph structure provided a more significant improvement. GraphSAGE unexpectedly achieved the highest accuracy (50.00\%), notably outperforming all non-graph methods, including activity-specific models like NeuPRINT (45.57\%) and LOLCAT (39.00\%). This suggests that when the node feature itself is weak and noisy (as shown in Section 3.1.3), aggregating information from neighbors via message passing provides crucial contextual information that aids classification. However, it is critical to note that even with this graph-based improvement, the ISI feature's top performance remains significantly worse than that of the other two features.

\noindent\textbf{Non-Graph Methods. } LOLCAT (57.00\%) achieved competitive scores on Spatial Feature. On the Neuronal Activity Feature, models designed for this data type showed moderate performance such as NeuPRINT (45.57\%). The poor performance of LOLCAT(39.00\%), a model designed for ISI distributions, reinforces our conclusion from Section 3.1.3: the low-resolution data and resulting event sparsity likely prevent the model from learning informative temporal patterns.

\section{Conclusion, Limitation and Future Works}
\label{sec:Conclusion}

This study benchmarked graph and non-graph methods for \textit{C. elegans} neuron classification using Spatial, Connection, and Neuronal Activity features. We found that GNNs, particularly GraphTransformer and GAT, significantly outperformed non-graph baselines on the Spatial and Connection features. These two features were found to be highly informative and significantly correlated. Conversely, the Neuronal Activity feature yielded poor performance across all models, which may be attributed to the data's low temporal resolution.

Our study has several limitations. First, our models were trained on a static functional connectome, whereas brain activity is inherently dynamic. Second, although we found Spatial and Connection features to be correlated, we did not explore their combination. Future work could explore dynamic GNN models to capture the temporal evolution properties of the network. Additionally, future research could investigate feature fusion (e.g., combining Spatial and Connection features) to examine potential gains.

\section{Compliance with ethical standards}
This research study was conducted retrospectively using animal subject (C. elegans) data made available in open access by Randi et al.~\cite{Randi}. Ethical approval was not required as confirmed by the licenses attached with the open access datasets.

\section{Acknowledgments}
\label{sec:acknowledgments}
This research was partially supported by the internal funds and GPU servers provided by the Computer Science Department of Emory University, the US National Science Foundation under Award Numbers 2312502 and 2319449. LM was not supported by any funds from US.

\bibliographystyle{IEEEbib}
\bibliography{ref}

\end{document}